\documentclass[jrfm,article,accept,pdftex,moreauthors]{Definitions/mdpi} 

\usepackage{graphicx}
\usepackage{lscape}
\usepackage{pdflscape}
\usepackage{booktabs}
\usepackage[labelformat=simple]{subcaption}

\DeclareCaptionLabelFormat{subcaptionlabel}{\normalfont(\textbf{#2}\normalfont)}
\firstpage{1} 
\pubvolume{1}
\issuenum{1}
\articlenumber{0}
\pubyear{2023}
\copyrightyear{2022}
\externaleditor{Academic Editor: {Zhaojun Yang and Thanasis Stengos}} 
\datereceived{13 November 2022} 
\daterevised{22 December 2022}
\dateaccepted{28 December 2022} 
\datepublished{} 
\hreflink{https://doi.org/}

\Title{The Effect of COVID-19 on Cryptocurrencies and the Stock Market Volatility: A Two-Stage DCC-EGARCH Model Analysis}

\TitleCitation{The Effect of COVID-19 on Cryptocurrencies and the Stock Market Volatility: A Two-Stage DCC-EGARCH Model Analysis}

\Author{{Apostolos Ampountolas} $^{1,{2}}$ \href{https://orcid.org/0000-0003-3992-6663}{\orcidicon} 
}

\AuthorNames{Apostolos Ampountolas}

\AuthorCitation{Ampountolas, Apostolos}

\address{%
$^{1}$ \quad {School of Hospitality Administration, {Boston} University, Boston, MA 02215, USA; aampount@bu.edu}\\ 
$^{2}$ \quad {Department of Mathematics,  College of Engineering, Design and Physical Sciences, \mbox{Brunel University London}, {Uxbridge} UB8 3PH, UK; apostolos.ampountolas@brunel.ac.uk;} 
\\
}

\abstract{This research examines the correlations between the return volatility of cryptocurrencies, global stock market indices, and the spillover effects of the COVID-19 pandemic. For this purpose, we employed a two-stage multivariate volatility exponential GARCH (EGARCH) model with an integrated dynamic conditional correlation (DCC) approach to measure the impact on the financial portfolio returns from 2019 to 2020. Moreover, we used value-at-risk (VaR) and value-at-risk measurements based on the Cornish--Fisher expansion (CFVaR). The empirical results show significant long- and short-term spillover effects. The two-stage multivariate EGARCH model's results show that the conditional volatilities of both asset portfolios surge more after positive news and respond well to previous shocks. As a result, financial assets have low unconditional volatility and the lowest risk when there are no external interruptions. Despite the financial assets' sensitivity to shocks, they exhibit some resistance to fluctuations in market confidence. The VaR performance comparison results with the assets portfolios differ. During the COVID-19 outbreak, the Dow (DJI) index reports VaR's highest loss, followed by the S\&P500. Conversely, the CFVaR reports negative risk results for the entire cryptocurrency portfolio during the pandemic, except for the Ethereum (ETH).}


\begin{document}

\section{Introduction}
Though the pandemic is still active and has uncertain long-term outcomes \citep{fisher2021impact}, at the same time, the US administration has lately announced that the pandemic is over \citep{NYT2022COVID}, probably because we have reached the stage of the COVID-19 outbreak known as endemic when the virus is ubiquitous but substantially less deadly than it was in 2020. Regrettably, the severe mortality toll from the COVID-19 epidemic is still being reported. Similarly, we are all also concerned about how this crisis affected the financial markets \citep{ashraf2020economic} and, more generally, the entire economic environment, including consumer behavior and intentions \citep{watson2021will, goldstein2021COVID}. From this perspective, asset management has already become essential for organizations in managing their assets to generate the highest returns due to today's difficult circumstances and highly volatile markets.

Asset allocation attempts to establish an equilibrium between risk and return by changing the ratio of each asset in a portfolio to achieve the goals, objectives, return expectations, and risk tolerance following an investment period for the investor. However, it is practical to anticipate that the topic of the COVID-19 pandemic and financial market volatility have, and will form, a significant interest in academic research in the near future. Future research topics might examine the effects of earlier occurrences resembling COVID-19, how COVID-19 may differ from those earlier events, and obtain an optimal portfolio in a highly dependent volatile financial market environment. Furthermore, along with studying the effects of previous pandemics, recent academic studies (see, \cite{MAZUR2021101690, alqaralleh2021evidence, akhtaruzzaman2022systemic, uddin2022stock}) also ominously predict significant events like COVID-19 and its economic repercussions.

Financial market information has become more widely available and has enhanced the relationship between market volatility and other factors. Based on price movements in other markets, investors forecast price changes. In other words, when one market's return grows, the other markets' returns may also alter simultaneously, forming a spillover effect. Therefore, Value-at-Risk (VaR) is one of the most commonly employed techniques for assessing market risk, a standardized volatility measurement tool \citep{doi:10.2469/faj.v52.n6.2039}. Moreover, Dynamic Conditional Correlation (DCC) is thus proven to be an effective and practical methodology for more robust market volatility and yielding conditional variances \mbox{\citep{tse2002multivariate}}. As such, this study investigates the dynamic correlation between four stock market indices and four different cryptocurrencies using data from January 2019 to December 2020. 

{The COVID-19 pandemic has significantly impacted the financial markets worldwide. In addition, the outbreak has had a particularly detrimental effect on cryptocurrencies' potential as alternative investments. This research adds to the growing literature by examining the connections between the financial assets of each portfolio and the respective substantial volatility dynamics. As such, the study extends the literature by examining the correlations between the return-volatility of cryptocurrencies and global stock markets indices, such as the S\&P500, DJI, GDAXI, and FTSE, the return-volatility spillover between cryptocurrencies, namely the Bitcoin, Ethereum, Cardano, and Ripple, and the effects of the COVID-19 pandemic on return-volatility by covering distinct peaks of the market during the pandemic. Considering the potential downside risk of investing in these two financial portfolios that are expanding simultaneously is critical. Therefore, studying the dynamics of the financial and cryptocurrencies bear markets during COVID-19 offers an unprecedented opportunity to examine. Comparing the behavior of cryptocurrencies to major stock market indices is worthwhile.} \cite{conlon2020safe} {findings show that Bitcoin does not function as a safe place when considering the impact on an S\&P500 portfolio that is balanced and includes exposure to Bitcoin. They found that the S\&P500 and Bitcoin trade is in perfect synchronization during the period under review, increasing the downside risk for an investor who has allocated money to Bitcoin.}

Moreover, unlike earlier studies that used GARCH, and other mean-variance methods, this study addresses the cross-asset return and conditional volatilities using a multivariate two-stage dynamic conditional correlation model, the DCC-EGARCH model. The DCC model adopts a conditional correlational and time-varying impact to properly evaluate the dynamic correlation structure for addressing the volatilities and estimated returns. In addition, the study measures the market risk using the well-known VaR model. It employs a four-moment modified VaR based on the Cornish--Fisher (CFVaR) expansion, which in addition to mean and variance, also considers skewness and Kurtosis and is considered more accurate than the two moments VaR \citep{favre2002mean, conlon2020safe, ali2021downside}. Because these allow us to observe and quantify how returns and volatility vary across markets, these approaches are preferable to the traditional time series~approach.

{This research analysis determines the essential positive and negative effects on the financial markets and quantifies the pandemic's impact. Moreover, the present research provides explanations of the market's volatility so that investor can diversify their investment approaches.}

The research is structured as follows: we start with a brief literature review in Section~\ref{sec2}, then in Section~\ref{sec3}, we describe the employed models, followed by defining the data collection in Section~\ref{sec4}, analyze the findings in Section~\ref{sec5}, and finally proceed to the study conclusion in Section~\ref{sec6}. 

\section{Research Background}\label{sec2}

The COVID-19 outbreak and the uncertainty arising from each country's administration restrictions have shown a growing number of academic literature discussing the implications of cryptocurrencies, market indices, and alternative investments, see, (\cite{conlon2020safe, alqaralleh2021evidence, goldstein2021COVID, MAZUR2021101690, nguyen2022correlation, uddin2022stock, yan2022garch}). For example, the emergence of COVID-19 has raised essential concerns about how the interaction between cryptocurrencies and other alternative investments has changed throughout the pandemic. According to \cite{conlon2020safe}, Bitcoin increased portfolio risk during high uncertainty, indicating it would not be a safe haven for investments. Similarly, \cite{nguyen2022correlation} claimed that the returns on the S\&P500 considerably influenced the returns on Bitcoin at times of high uncertainty. During COVID-19 and other volatile times, stock market shocks impacted Bitcoin's volatility. \cite{uddin2022stock} study investigated the interconnected dynamics of the affected Asian and worldwide financial markets in response to the coronavirus (COVID-19) pandemic epidemic. Their results showed that the COVID-19 outbreak had caused a significant, positive reliance among the markets and an enhanced tendency for co-movements across the upper time horizon. \cite{yan2022garch} study examined the dynamic conditional correlations between 10 cryptocurrencies throughout the COVID-19-affected timeframe of 2017 to 2022. According to their findings, all cryptocurrency return growth rates increased from the pre-COVID-19 period to the COVID-19 period, and COVID-19 had a favorable impact on cryptocurrency returns. From pre-COVID-19 to COVID-19, the average dynamic correlations between the return indices of Bitcoin, Ethereum, and other cryptocurrencies were very high. {It is obvious that investments in cryptocurrencies are frequently seen with high volatility and risk. In their work,} \cite{almeida2022uncertainty} {examined a sample of seven cryptocurrencies, including the period of the COVID-19 pandemic employing value-at-risk (VAR) and conditional VAR to quantify risk. The findings show that, other than the Tether, the cryptocurrencies displayed comparable patterns of risk and uncertainty. Moreover,} \cite{umar2020time} {used the wavelet method to investigate how COVID-19 has affected the volatility of significant fiat and cryptocurrency markets. The findings for each index combination in the study are relatively consistent and support the hypothesis that cross-currency hedge techniques, which could be successful in a market under normal conditions, are more likely to fail in market shocks such as the COVID-19~outbreak.}

\section{Methodology}\label{sec3}

This section describes the proposed two stages GARCH modeling framework. To simulate the time-varying volatility in the stock market indices and the cryptocurrency return series, we first employed the alternative Generalized Autoregressive Conditionally Heteroscedastic (GARCH)-type specification, the exponential GARCH (EGARCH) model. 
Second, we follow \cite{conlon2020safe} and estimate downside risk---value-at-risk (VaR), also based on Cornish--Fisher expansion (CFVaR) for each financial portfolio in this study. Finally, the section describes the criteria for choosing the best GARCH-type specifications.

\subsection{Multivariate GARCH}
The GARCH family of models introduced by \cite{bollerslev1986generalized} has been extensively employed in financial modeling. The generalized autoregressive conditional heteroscedasticity model (GARCH) extends the autoregressive conditional heteroscedasticity (ARCH) model. The GARCH model maintains the unconditional variance constant while enabling the conditional variance to vary over time due to prior errors \citep{bollerslev1986generalized}. Due to the significant modifications made to the original GARCH model, GARCH models have been used in this context to forecast the variance of time series, not only in the financial sector to estimate stock volatility performance but also to estimate the volatility of cryptocurrencies or other financial instruments (\cite{trucios2019forecasting, urquhart2019bitcoin, caporale2019day, conlon2020safe, alqaralleh2021evidence, ampountolas2022cryptocurrencies}), and to assess the accuracy of forecasts. Moreover, GARCH models are employed except for assessing financial instruments and evaluating the forecasting performance of time series datasets in other industries \citep{forecast3030037}.

The current study will employ a multivariate GARCH model in two stages to capture the asymmetric dependence structure \citep{ghalanos2015rmgarch}. In the first stage, a univariate GARCH process is established, together with the limitations for each series. We use the EGARCH (1, 1) model to analyze the heteroscedasticity in the daily returns of portfolio instruments. To appropriately represent the leverage effect, we are utilizing the well-known EGARCH model, which typically provides better fits than the traditional GARCH model. It also includes an additional asymmetric term. The model's estimation and evaluation are tasks for the second stage. Since it permits the conditional correlation matrix to be time-dependent, ensuring that it is positive definite and practical when modeling high-dimensional data sets, the dynamic conditional correlation (DCC-GARCH) method is commonly used in practice, as per \cite{bauwens2006multivariate}. Furthermore, we also discuss the conditional variance distribution's form parameter, which denotes excessive kurtosis in the studied series. 

\subsubsection{EGARCH Model}
The volatility of financial instruments and commodities is frequently modeled using the GARCH-class models. A maximum likelihood (ML) estimator is practically used to estimate the GARCH models' vector of unstructured parameters. The asymmetric relationship between asset returns and volatility changes, which is crucial when working with time series financial data, is ignored by the symmetric ARCH and GARCH models, which can capture volatility clustering and leptokurtosis.  \cite{nelson1991conditional} introduced the exponential generalized autoregressive conditional heteroscedasticity (EGARCH) model, a rigorous parametric form, to solve the drawbacks of symmetric models, such as the leverage effect conditional heteroscedasticity. As such, he achieved a non-negative conclusion by using the natural logarithm of the conditional variance. The EGARCH logarithm is described as follows: 
\begin{equation}
	\log(h_{t})= \alpha_0 +\alpha_1 \Bigg(\Bigg| \frac{\epsilon_{t-1}}{\sqrt{h_{t-1}}} \Bigg| - \mathbb{E} \Bigg|\frac{\epsilon_{t-1}}{\sqrt{h_{t-1}}} \Bigg|\Bigg) +\xi \frac{\epsilon_{t-1}}{\sqrt{h_{t-1}}} +\gamma_1\log (h_{t-1})
\end{equation}
where $\alpha_1 \Bigg(\Bigg| \frac{\epsilon_{t-1}}{\sqrt{h_{t-1}}} \Bigg| - \mathbb{E} \Bigg|\frac{\epsilon_{t-1}}{\sqrt{h_{t-1}}} \Bigg|\Bigg)$ denotes the magnitude effect and $\xi \frac{\epsilon_{t-1}}{\sqrt{h_{t-1}}}$ the sign effect. Therefore, if $\xi  < 0$, negative innovations (negative news) increase volatility more than positive innovations of equal size. The volatility specification's logarithmic adjustment suggests that this model's parameters are not restricted to positive values. The parameter for conditional variance $h_t$ is the GARCH coefficient $\gamma_1$ is typically close to 1, and the closer, the higher the conditional volatility.

\subsubsection{DCC-GARCH Model} 
\cite{engle2002dynamic} proposed the DCC-GARCH model, which features a dynamic conditional correlation structure. The correlations are first estimated in the DCC-GARCH model, then the GARCH parameters. Because $H_t$ represents the conditional covariance matrix and the conditional variances must also agree, the dynamic correlation model allows $R$ to vary over time, and its parameterizations of $R$ exactly satisfy all specifications. The matrix $R_t$ is the conditional correlation matrix, while the diagonal matrix $D_t$ has time-varying parameterizations \citep{engle2002dynamic}. The conditional correlation variance of the DCC-GARCH is parameterized as follows, where $\rho_{ij,t}$ stands for the correlation between the $i$th and $j$th return series. Therefore, \cite{engle2002dynamic} is regarded as a dynamic matrix process that may be written as
\begin{equation}
\boldsymbol{H}_{t}=\boldsymbol{D}_{t} \boldsymbol{R}_{t} \boldsymbol{D}_{t}
\end{equation}
\begin{equation}
	\boldsymbol{D}_{t} = \text{diag} \Big\{D_{1t}, D_{2t}, \dots, D_{kt}   \Big\} \quad \boldsymbol{R}_{t} = \Big[ \rho_{ij,t} \Big]
\end{equation}
\begin{equation}
\begin{split}
\boldsymbol{Q}_{t}= \boldsymbol{\bar{Q}} + \alpha\left(z_{t-1} z_{t-1}^{\prime} - \boldsymbol{\bar{Q}}\right) + \beta\left(\boldsymbol{Q}_{t-1} - \boldsymbol{\bar{Q}}\right)  \\	
 = \boldsymbol{\bar{Q}}(1-\alpha-\beta)+\alpha z_{t-1} z_{t-1}^{\prime}+\beta \boldsymbol{Q}_{t-1}
\end{split}
\end{equation}
\begin{equation}
\label{dcc}
\boldsymbol{R}_{t}= diag\left(\boldsymbol{Q}_{t}\right)^{-1 / 2} \boldsymbol{Q}_{t}diag\left(\boldsymbol{Q}_{t}\right)^{-1 / 2}
\end{equation}
where $\alpha$ is a positive and $\beta$ non-negative scalar parameter such that $\alpha + \beta < 1$ to ensure the stationarity and positive definiteness of $\boldsymbol{Q_t}$, $\bar{Q}$ is the unconditional matrix of standardized errors ($z_t$), which is entered into the equation via the covariance targeted portion $\boldsymbol{\bar{Q}}(1-\alpha-\beta)$, and $\boldsymbol{Q_0}$ is positive definite \citep{bauwens2006multivariate, ghalanos2015rmgarch}, even though this method creates reliable correlation matrices, it does not give positive definiteness; as a result, the correlation matrix $\boldsymbol{R}$ is created by rescaling $\boldsymbol{Q_t}$ as in Equation (\ref{dcc}) \mbox{\citep{bauwens2006multivariate}.} 

\subsection{Value-at-Risk (VaR)---Downside Risk}
Value at Risk (VaR) is a statistical tool used in assessing the volatility of stock indexes and has gained popularity as a tool for market risk analysis \citep{favre2002mean}. It quantifies the highest loss financial assets may sustain at a specific time under a particular confidence level. If returns are normally distributed, a two-moment Value-at-Risk (VaR) that considers return and asset standard deviation may be used to calculate the downside risk. 
 VaR is defined as:
\begin{equation}
	VaR_{p(\alpha)} = -(\mu_p + Z_a\sigma_p) 
\end{equation}
where $\mu_p$ denotes the mean of the portfolio daily returns, and $\sigma_p$ the standard deviation, respectively. Moreover, $Z_{\alpha}$ refers to the $\alpha$ quantile for a standard normal distribution. In general, the typical values for $(1-\alpha)$ are 90\%, 95\%, and 99\% which we also report in this study to indicate the differences.

In periods of market instability when returns are not normally distributed, a four-moment modified VaR, which supplements skewness and excess kurtosis to the two-moment VaR, gives more robust estimates on the downside risk of a portfolio. The downside risk is then quantified using higher statistical moments when simulating potential diversification advantages across various financial assets under consideration. 

Using the VaR calculation will result in skewed findings since the log returns of financial assets are frequently skewed and, as a result, not normally distributed. One workable alternative is using the Cornish--Fisher expansion \citep{cornish1938moments} to calculate the quantiles of such a non-normal distribution \citep{favre2002mean}. \cite{conlon2020safe}, and \cite{ali2021downside} used a similar process in measuring the downside risk in cryptocurrencies and the Dow Jones Islamic stock indices, respectively. 

The following equation describes how the Cornish--Fisher expansion, employing four moments, converts a conventional Gaussian variable z into a non-Gaussian random variable~\emph{Z}: 
\begin{equation}\label{eq:CF}
Z_{CF}=z_c+(z^2_c-1)\frac{S}{6} + (z^3_c-3z_c)\frac{K}{24} - (2z^3_c-5z_c)\frac{S^2}{36},
\end{equation}
where $Z_c$ as the critical value for probability (1 $-$ $\alpha$), $S$ as a skewness parameter, and $K$ as an excess kurtosis parameter. It is important to realize that $S$ and $K$ are parameters. As such, they can be very different from the actual skew and excess kurtosis of the obtained distribution following the Cornish--Fisher expansion.
\begin{equation}\label{eq:VaR}
VaR = W(\mu - Z_{CF} \sigma),
\end{equation}
where $W$ is the amount at risk or portfolio, and $\sigma$ refers to the yearly standard deviation.

\subsection{Model Selection}
To investigate the dynamic correlation between equity and commodity markets, we rely on DCC- GARCH.
The maximum likelihood (MLE) estimation was used to estimate the (($p$ = 1) and ($q$ = 1)) GARCH models for this study. According to \cite{hansen2005forecast}, there was little evidence that alternative, more complex models could outperform a GARCH(1,1). A GARCH(1,1) model with constant mean could be expressed as follows: $y_t =\mu_t +e_t$ with $e_t\sim N(0,\sigma^2_t)$. 
\begin{equation}
\begin{cases}
		y_t =\mu_t +e_t, \\ 
		e_t = \sqrt{h_t} \cdot \eta_{t}, \eta_t \sim iid(0,1),\\ 
		h_t = \alpha_0 + \alpha_1\epsilon^2_{t-1}+\gamma_1 h_{t-1}
\end{cases}
\end{equation}

The Akaike information criterion (AIC) \citep{akaike1981likelihood} was used to assess the discrimination as follows:
\begin{equation} \label{akaike}
	\text{AIC} = 2k - 2\ln L\left(\hat{\Theta}\right)\text{,}
\end{equation}
where $k$ represents the total number of unknown parameters, $\Theta$ is the vector of unknown parameters, and $L(\hat{\Theta})$ their maximum likelihood (ML) estimations; therefore, the optimal models were obtained by eliminating the criteria. Additionally, AIC promotes more sophisticated models in general \citep{gilli2019numerical}. 
  
\section{Data}\label{sec4}
{We examine the impact of the COVID-19 pandemic taking into account two portfolios, including major financial assets' daily close prices not only in the US market but also in Europe:} (a) four stock market indices, namely in the US: the S\&P 500 (GPSC), Dow Jones (DJIA), and Europe, the German DAX Performance (GDAXI), and in the UK the FTSE 100 (FTSE); (b) daily historical closing prices of cryptocurrencies such as the Bitcoin (BTC-USD), Ethereum (ETH-USD), Ripple (XRP-USD), and Cardano (ADA-USD) and correlates them with the COVID-19 period. The data set observations covering the period from 1 January 2019 to  31 December 2020 ({1 January 2019} 
 0:00 to {31 December 2020} 23:59 EDT), and the entries for each observation included the exchange date, time, symbol, open, high, low, close price, and volume for all trades in US dollars. {The data set range spans the coronavirus outbreak period after a significant upswing and downswing in asset prices following the second COVID-19 wave.} The data sets were obtained from Alpha Vantage for the stock market indices and from ``\url{CryptoDataDownload.com} (accessed on 14 June 2022),\endnote{Data are publicly available at {\url{https://www.cryptodatadownload.com/data/}} (accessed on 14 June 2022).}'' for the cryptocurrencies in the EDT time zone.

Portfolio returns were calculated using the formula $r_{t} = ln\bigg(\frac{P_t}{P_{t-1}}\bigg)$ and define\linebreak $\alpha_t = r_t - E_{t-1}$, where $P_t$ was the closing price of the assets on time $t$. 
We initiated our research by determining the data stationarity using the Dickey--Fuller (ADF) coefficient test for identifying non-seasonal unit roots \citep{dickey1979distribution}. Thus, as the evaluated models needed stationary data, we computed the log returns and performed the augmented Dickey--Fuller (ADF) test to confirm that there were no unit roots. The unit root tests rejected the null hypothesis at 0.01 level for the analyzed portfolio returns. We also used the KPSS stationarity test \citep{kwiatkowski1992testing}, which accepts the null hypothesis for portfolio return rates, to confirm the findings of the ADF test ({Table}
~\ref{tab:markets_corr}---stock market indices, {Table}~\ref{tab:crypto_corr}---cryptocurrencies assets, respectively).

The models are arranged in a certain order using the Akaike Information Criterion (AIC). The Jarque--Bera Test is a test to determine if a set of data values follows the normal distribution based on the data's skewness and kurtosis \citep{jarque1987test}. The test statistic equation incorporating skewness and kurtosis is: $JB=\frac{n}{6}\left(S^2+\frac{1}{4}(K-3)^2\right)$ where n = the number of values for the data. $S$ is the sample skewness (how much the data lean away from the mean), and $K$ refers to the sample kurtosis (how wide the distribution's tails~are).

\begin{table}[H]
\caption{Correlation and Unit Root Test---Stock market indices.}
\label{tab:markets_corr}
\renewcommand{\arraystretch}{1.05}\setlength{\tabcolsep}{2.6mm}
\begin{tabular}{p{0.13\linewidth}rrrrp{0.1\linewidth}rr}
\toprule
\multicolumn{1}{c}{} & \multicolumn{1}{c}{\textbf{DJI}} & \multicolumn{1}{c}{\textbf{FTSE}} & \multicolumn{1}{c}{\textbf{GDAXI}} & \multicolumn{1}{c}{\textbf{GSPC}} & \multicolumn{1}{c}{} & \multicolumn{1}{c}{\textbf{B}} & \multicolumn{1}{c}{\textbf{KPSS}} \\
\midrule
DJI                  & 1                       &                          &                           &                          &                      & $-$6.4190 *                & 0.06104                  \\
FTSE                 & 0.704                  & 1                        &                           &                          &                      & $-$7.1827 *                & 0.07928                  \\
GDAXI                & 0.693                  & 0.869                   & 1                         &                          &                      & $-$7.0149 *                & 0.05182                  \\
GSPC                 & 0.980                  & 0.681                   & 0.677                    & 1                        &                      & $-$6.0856 *                & 0.06295    \\
\bottomrule             
\end{tabular}

\noindent\footnotesize{Note: * Significant at the 0.05 level; ADF: augmented Dickey--Fuller statistics; KPSS: KPSS test statistics using residuals from regressions.}
\end{table}

\vspace{-12pt}

\begin{table}[H]
\caption{Correlation and Unit Root Test---Cryptocurrencies.}
\label{tab:crypto_corr}
\renewcommand{\arraystretch}{1.05}\setlength{\tabcolsep}{3mm}
\begin{tabular}{p{0.13\linewidth}rrrrp{0.1\linewidth}rr}
\toprule
\multicolumn{1}{c}{} & \multicolumn{1}{c}{\textbf{BTC}} & \multicolumn{1}{c}{\textbf{ETH}} & \multicolumn{1}{c}{\textbf{XRP}} & \multicolumn{1}{c}{\textbf{ADA}} & \multicolumn{1}{c}{} & \multicolumn{1}{c}{\textbf{B}} & \multicolumn{1}{c}{\textbf{KPSS}} \\
\midrule
BTC                  & 1                       &                         &                         &                         &                      & $-$8.6000 *                   & 0.20587                  \\
ETH                  & 0.834                 & 1                       &                         &                         &                      & $-$8.6093 *                & 0.16746                  \\
XRP                  & 0.554                 & 0.616                 & 1                       &                         &                      & $-$7.7385 *                & 0.04243                  \\
ADA                  & 0.681                 & 0.724                 & 0.716                 & 1                       &                      & $-$8.6460 *                & 0.11464   \\
\bottomrule              
\end{tabular}

\noindent\footnotesize{Note: * Significant at the 0.05 level; ADF: augmented Dickey--Fuller statistics; KPSS: KPSS test statistics using residuals from regressions.}
\end{table}

\begin{figure}[H]
\includegraphics[width=0.98\linewidth, height=0.45\textheight]{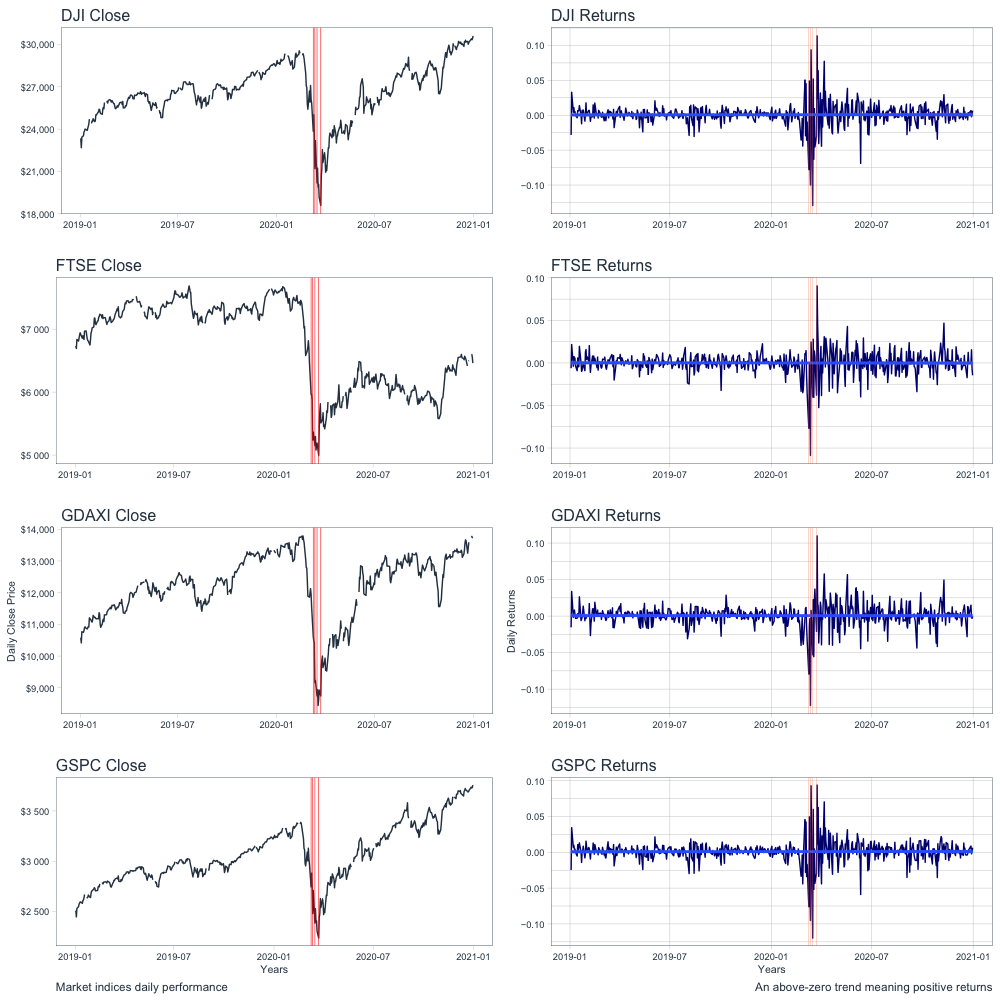}
\caption{{{Stock} 
 market indices performance and returns period January 2019--December 2020.}}
\label{fig:stock_prices}
\end{figure}

Figure \ref{fig:stock_prices} presents the historical price performance for the stock market indices and the log returns during the observed period. The red lines indicate that the market declined significantly during the four days in March 2020, generating excessive losses. Similarly, Figure \ref{fig:stock_prices} suggests that the log returns were moderately symmetrically distributed before the COVID-19 outbreak, with some peaks during the observed period. The stock market indices had experienced growth until the end of February or the beginning of March when the first actions in response to the coronavirus pandemic were disclosed.

Table~\ref{summarystock} illustrates the summary statistics for the stock market portfolio returns. Each stock market index group in this study has 516 daily portfolio returns. To gain a more transparent overview of the data, we separated the dataset into three periods, i.e., the entire dataset period 2019 to 2020, one year before the coronavirus outbreak (2019), and one year during the heavy coronavirus pandemic period (2020) (Table \ref{summarystock}). We note that the mean daily returns are positive for every assessed period for each market index and significantly during the pandemic period, respectively, except for the FTSE index, which reports a negative mean during the pandemic outbreak. On one side, it was expected as the European market indices have strongly crashed due to inconsistent decisions taken by the countries' governments and the imposed restrictions for the pandemic. Thus, in Table \ref{summarystock}, the statistics provide such observations as the standard deviation that evaluates the market volatility and how widely prices are dispersed from the average price. Therefore, we observed that the standard deviations were higher for the DJI index than any other market indices returns for the entire study period and during the pandemic, followed by the S\&P500. It confirms the fact that during the COVID outbreak between 12 February and 23 March  2020, the Dow Industrial Index (DJI) lost 37\% of its value in four trading days \citep{forbes2021}. The market returns indicate a significant frequency of negative returns showing that the distribution has a left-tail skewness in addition to the Sharpe ratio. 

Furthermore, we have performed and provided in the tables the Jarque--Bera (J\&B) test for all data frequencies, a formal test for the standard series distribution. The time series across all stock market indices groups are not distributed normally, according to the Jarque--Bera (J\&B) test results for normality. At the 1\% level of statistical significance, these findings reject the null hypothesis of a unit root and constant variance.  
\begin{table}[H]
\caption{Descriptive statistics---Stock market indices returns January 2019--December 2020.}
\label{summarystock}
\resizebox{\textwidth}{!}{%
\begin{tabular}{lrrrrrrrrrr}
\toprule
\multicolumn{1}{c}{} & \multicolumn{1}{c}{\textbf{Mean}} & \multicolumn{1}{c}{\textbf{Std}} & \multicolumn{1}{c}{\textbf{Min}} & \multicolumn{1}{c}{\textbf{Max}} & \multicolumn{1}{c}{\textbf{Kurt}} & \multicolumn{1}{c}{\textbf{Skew}} & \multicolumn{1}{c}{\textbf{Q0.25}} & \multicolumn{1}{c}{\textbf{Q0.75}} & \multicolumn{1}{c}{\textbf{SR}} & \multicolumn{1}{c}{\textbf{J\&B Test}} \\
\midrule
\multicolumn{5}{l}{\textbf{Panel A: (1 January 2019--31 December 2020)}}                                                               & \textbf{}                & \textbf{}                & \textbf{}                 & \textbf{}                 & \textbf{}              & \textbf{}                     \\
DJI                  & 0.00067                  & 0.01710                 & $-$0.12927                & 0.11365                 & 16.16261                 & $-$0.61453                 & $-$0.00415                  & 0.00657                   & $-$0.05334               & 5532.72   ***                      \\
FTSE                 & 0.00002                  & 0.01392                 & $-$0.10874                & 0.09053                 & 12.78455                 & $-$0.95778                 & $-$0.00533                  & 0.00624                   & $-$0.06849               & 3518.59   ***                      \\
GDAXI                & 0.00063                  & 0.01580                 & $-$0.12239                & 0.10976                 & 13.42626                 & $-$0.69903                 & $-$0.00464                  & 0.00733                   & $-$0.05790               & 3836.29   ***                      \\
GSPC                 & 0.00091                  & 0.01613                 & $-$0.11984                & 0.09383                 & 14.63959                 & $-$0.69860                 & $-$0.00358                  & 0.00724                   & $-$0.05559               & 4553.70   ***                      \\\midrule
\multicolumn{5}{l}{\textbf{Panel B: (1 January--31 December 2019)}}                                                               & \textbf{}                & \textbf{}                & \textbf{}                 & \textbf{}                 & \textbf{}              & \textbf{}                     \\
DJI                  & 0.00081                  & 0.00776                 & $-$0.03046                & 0.03292                 & 3.39944                  & $-$0.60943                 & $-$0.00264                  & 0.00514                   & $-$0.16717               & 133.07   ***                       \\
FTSE                 & 0.00047                  & 0.00733                 & $-$0.03231                & 0.02253                 & 2.22334                  & $-$0.39732                 & $-$0.00420                  & 0.00525                   & $-$0.17988               & 56.51   ***                        \\
GDAXI                & 0.00091                  & 0.00871                 & $-$0.03107                & 0.03370                 & 2.09444                  & $-$0.30200                 & $-$0.00331                  & 0.00577                   & $-$0.14812               & 48.02   ***                        \\
GSPC                 & 0.00101                  & 0.00778                 & $-$0.02978                & 0.03434                 & 3.40344                  & $-$0.57982                 & $-$0.00227                  & 0.00576                   & $-$0.16505               & 131.86   ***                        \\\midrule
\multicolumn{5}{l}{\textbf{Panel C: (1 January--31 December 2020)}}                                                               & \textbf{}                & \textbf{}                & \textbf{}                 & \textbf{}                 & \textbf{}              & \textbf{}                     \\
DJI                  & 0.00053                  & 0.02288                 & $-$0.12927                & 0.11365                 & 8.93517                  & $-$0.47372                 & $-$0.00588                  & 0.00898                   & $-$0.02303               & 834.02   ***                       \\
FTSE                 & $-$0.00043                 & 0.01825                 & $-$0.10874                & 0.09053                 & 7.55239                  & $-$0.76400                 & $-$0.00759                  & 0.00970                   & $-$0.03221               & 613.35   ***                       \\
GDAXI                & 0.00035                  & 0.02056                 & $-$0.12239                & 0.10976                 & 8.34480                  & $-$0.57867                 & $-$0.00685                  & 0.00993                   & $-$0.02621               & 733.12   ***                       \\
GSPC                 & 0.00081                  & 0.02144                 & $-$0.11984                & 0.09383                 & 8.28656                  & $-$0.55716                 & $-$0.00565                  & 0.00906                   & $-$0.02377               & 722.05   ***     \\
\bottomrule                 
\end{tabular}%
}
\noindent\footnotesize{Note:   *** Significant at the 0.001 level; SR: Sharpe Ratio; J\&B: Jacque--Bera test statistics.}
\end{table}

Table \ref{tab:markets_corr} presents the Pearson correlation coefficients for the stock market indices. As such, the S\&P500 index (GSPC) reports a high correlation with the Dow Jones index (DJI) (0.980), and similarly, the German DAX index (GDAXI) with the Financial Times index (FTSE) (0.869), respectively. Moreover, the S\&P500 index presents a moderate correlation with the FTSE index (0.681) and the GDAXI (0.677). Table \ref{tab:markets_corr} also shows the Augmented Dickey--Fuller (ADF) \citep{dickey1979distribution} and the KPSS \citep{kwiatkowski1992testing} tests, which could indicate whether the data series are stationary. The KPSS test is frequently employed in empirical studies to examine trend stationarity. It serves as an addition to the traditional ADF unit root test when reviewing the characteristics of time series data. 

\begin{figure}[H]
\includegraphics[width=0.98\linewidth, height=0.45\textheight]{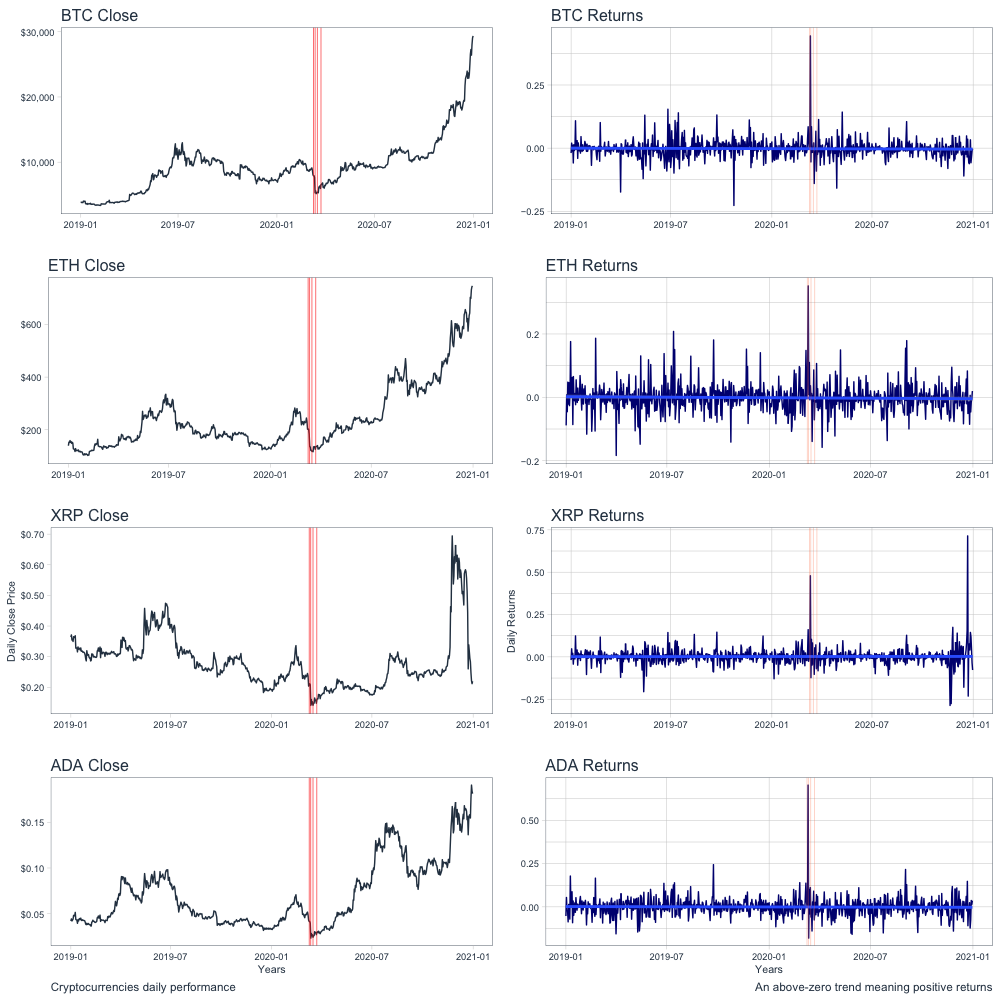}
\caption{{{Cryptocurrencies} 
 price-performance and returns period January 2019--December 2020.}}
\label{fig:crypto_returns}
\end{figure}

Figure \ref{fig:crypto_returns} displays this study's daily closing prices of the four cryptocurrencies. We observe that each cryptocurrency has a distinct tendency; for instance, Bitcoin and Ethereum followed a rising trend that held steady up to the days of the market crash in February and March, when the first coronavirus response measures were revealed. Similarly, we see that Cardano exhibits a nearly identical pattern through the end of January; as a result, they might be correlated. On the other hand, Ripple initially showed a stabilized gain before following a dramatic fall pattern until the beginning of April. Finally, the charts indicate that the prices of the remaining three cryptocurrencies have been steadily rising since the second half of April, except for Ripple, where we initially see a stable pattern followed by a sharp surge and decline. 

The daily log returns of the observed market price indices for all exchanges trading in the studied cryptocurrencies are also shown in Figure \ref{fig:crypto_returns}, jointly with their associated daily log returns. The charts exhibit that the log returns are moderately symmetrically distributed, with specific spikes within the analyzed period, comparable to the earlier data presented in the descriptive statistics table. Moreover, Ripple (XRP-USD) demonstrates the tenuous proof of volatility clustering, which, although symmetrically distributed, does not follow similar spillover effects as the other three cryptocurrencies.    

Table~\ref{tab:crypto_descriptive} reports the summary statistics for the cryptocurrency market returns. Similar to the stock market indices dataset, we have separated the statistics into three periods. For the entire period, the Bitcoin (BTC-USD) ($-$0.002) and the Ethereum (ETH-USD) ($-$0.001) present a negative mean (Panel A), however a year before the COVID-19 outbreak (Panel B), only the Bitcoin (BTC-USD) reports negative returns mean ($-$0.001). The remaining cryptocurrencies report positive means (Panel B). The standard deviation for Bitcoin (BTC-USD) 0.039 is the lowest among the cryptocurrencies, which indicates low volatility in each reported period, followed by Ripple (XRP-USD) 0.039 in the pre-COVID-19 pandemic period (Panel B). For the remaining cryptocurrencies, the standard deviations are 0.047, 0.055, and 0.057 for Ethereum (ETH-USD), Ripple (XRP-USD), and Cardano (ADA-USD), respectively (Panel A). The high level of volatility is characterized by comparatively high standard deviations as well as minimum and maximum rates. Additionally, the average return for all four cryptocurrencies is positive.
XRP is the only cryptocurrency that displays more consistent positive returns over the evaluation period, according to Table~\ref{tab:crypto_descriptive}. It implies that XRP has a lower systematic risk in the cryptocurrency market because its correlation with the market is more minimal than other cryptocurrencies. When the time correlation drops and a short uncorrelated It\^{o} process is revealed, the GARCH models typically exhibit weak persistent behavior, especially over extended time horizons \citep{carbone2004time}. As a result, investors looking to build a market portfolio might be more interested in it. Lastly, the Jarque--Bera (J\&B) test \citep{jarque1987test}  reveals that the price returns of all cryptocurrencies generally deviate from normality (Table \ref{tab:crypto_descriptive}).

Table \ref{tab:crypto_corr} shows that the cryptocurrency pairs have positive and significant Pearson correlation coefficients.  Additionally, a significant correlation of 0.834 is reported between Bitcoin and Ethereum. Furthermore, Cardano exhibits the lowest correlation with Bitcoin (0.681) and a moderately high correlation with Ethereum (0.724) and Ripple (0.716). Similarly, Ripple has the lowest correlation 
 0.554 and 0.616 with Bitcoin and Ethereum, respectively. Table \ref{tab:crypto_corr} also presents the unit root tests for the study's cryptocurrencies.

\begin{table}[H]
\caption{Descriptive Statistics---Cryptocurrencies returns January 2019--December 2020.}
\label{tab:crypto_descriptive}
\resizebox{\textwidth}{!}{%
\begin{tabular}{lrrrrrrrrr}
\toprule
\multicolumn{1}{c}{} & \multicolumn{1}{c}{\textbf{Mean}} & \multicolumn{1}{c}{\textbf{Std}} & \multicolumn{1}{c}{\textbf{Min}} & \multicolumn{1}{c}{\textbf{Max}} & \multicolumn{1}{c}{\textbf{Kurt}} & \multicolumn{1}{c}{\textbf{Skew}} & \multicolumn{1}{c}{\textbf{Q0.25}} & \multicolumn{1}{c}{\textbf{Q0.75}} & \multicolumn{1}{c}{\textbf{J\&B Test}} \\
\midrule
\multicolumn{5}{l}{\textbf{Panel A: (1 January 2019--31 December 2020)}}                                                                        &                          &                          &                           &                           &                               \\
Bitcoin              & $-$0.002                   & 0.039                   & $-$0.227                  & 0.444                   & 25.364                   & 1.914                    & $-$0.018                    & 0.012                     & 19,759.237   ***                  \\
Ethereum             & $-$0.001                   & 0.047                   & $-$0.183                  & 0.352                   & 6.541                    & 1.010                    & $-$0.024                    & 0.018                     & 1406.006   ***                   \\
Ripple               & 0.002                    & 0.055                   & $-$0.287                  & 0.714                   & 48.688                   & 3.474                    & $-$0.018                    & 0.019                     & 72,660.308   ***                  \\
Cardano              & 0.000                    & 0.057                   & $-$0.181                  & 0.704                   & 32.830                   & 2.711                    & $-$0.028                    & 0.027                     & 33,257.031   ***                  \\\midrule
\multicolumn{5}{l}{\textbf{Panel B: (1 January--31 December 2019)}}                                                                        &                          &                          &                           &                           &                               \\
Bitcoin              & $-$0.001                   & 0.039                   & $-$0.227                  & 0.154                   & 5.735                    & $-$0.160                   & $-$0.016                    & 0.015                     & 485.316   ***                    \\
Ethereum             & 0.001                    & 0.047                   & $-$0.183                  & 0.208                   & 3.979                    & 0.597                    & $-$0.021                    & 0.020                     & 253.799   ***                    \\
Ripple               & 0.002                    & 0.039                   & $-$0.206                  & 0.146                   & 4.101                    & $-$0.143                   & $-$0.014                    & 0.019                     & 248.119   ***                    \\
Cardano              & 0.002                    & 0.047                   & $-$0.156                  & 0.244                   & 2.988                    & 0.456                    & $-$0.023                    & 0.027                     & 143.205   ***                    \\\midrule
\multicolumn{5}{l}{\textbf{Panel C: (1 January--31 December 2020)}}                                                                        &                          &                          &                           &                           &                               \\
Bitcoin                  & $-$0.003                   & 0.040                   & $-$0.159                  & 0.444                   & 43.801                   & 3.839                    & $-$0.019                    & 0.010                     & 29,336.785   ***                  \\
Ethereum                  & $-$0.004                   & 0.048                   & $-$0.158                  & 0.352                   & 9.196                    & 1.412                    & $-$0.030                    & 0.016                     & 1370.999   ***                   \\
Ripple                  & 0.001                    & 0.067                   & $-$0.287                  & 0.714                   & 42.206                   & 3.806                    & $-$0.022                    & 0.019                     & 27,287.361   ***                  \\
Cardano                  & $-$0.003                   & 0.065                   & $-$0.181                  & 0.704                   & 37.407                   & 3.472                    & $-$0.035                    & 0.026                     & 21,473.441   ***     \\
\bottomrule            
\end{tabular}}
\noindent\footnotesize{Note:   *** Significant at the 0.001 level. J\&B: Jacque--Bera test statistics.}
\end{table}

\section{Results}\label{sec5}
This study employed an estimation approach in two steps for each portfolio. Constraints are established in the first stage for the univariate GARCH algorithm for each series. As such, the GARCH model with the maximum likelihood (MLE) estimator is frequently utilized since these features necessitate volatility modeling. Moreover, the exponential GARCH model (EGARCH), one of the well-known and often used model specifications for the GARCH process, offers better fits than traditional GARCH (1,1) models \citep{lee1994asymptotic}. Hence, the leverage impact is captured by an additional asymmetric term in the EGARCH model. As a result of comparing the stock market and cryptocurrencies portfolios, our objective supports employing multivariate GARCH rather than univariate~GARCH. 

We employed a multivariate DCC-GARCH model in the second step to estimate pairwise models between the study portfolios. According to statistically significant estimated coefficients that are primarily near the 1\% level, the volatility of an asset is commonly strongly influenced by its historical squared shocks and historical volatility, independent of the asset pair under study. We can observe evidence of considerable cross-market impacts between the variability of the returns of all assets pairs, particularly with shock and volatility spillovers, as demonstrated by the findings of the DCC-GARCH model in \mbox{Tables \ref{tab:dcc_stock} and \ref{tab:my-table}}. At the 1\% level, the estimates of $\alpha_{1}$, and $\beta_{1}$ are statistically significant. The portfolio pairs in the paired models exhibit bidirectional transmission and volatility relationships. The results indicate that lagged shocks and volatility strongly and significantly impact both the own volatility spillovers~($\beta_{1}$) and the conditional volatility that is currently existing in both assets portfolio~($\alpha_{1}$). 

\begin{table}[H]
\caption{Results from multivariate EGARCH (DCC) model---Stock Market Indices.}
\label{tab:dcc_stock}
\renewcommand{\arraystretch}{1.25}
\resizebox{\textwidth}{!}{%
\begin{tabular}{lp{0.153\linewidth}p{0.15\linewidth}p{0.15\linewidth}p{0.14\linewidth}p{0.15\linewidth}p{0.15\linewidth}p{0.1\linewidth}p{0.1\linewidth}}
\toprule
\multicolumn{1}{c}{} & \multicolumn{1}{c}{\boldmath{$\omega$}} & \multicolumn{1}{c}{\boldmath{$\alpha_1$}} & \multicolumn{1}{c}{\boldmath{$\beta_1$}} & \multicolumn{1}{c}{\boldmath{$\gamma_1$}} & \multicolumn{1}{c}{\textbf{Skew}} & \multicolumn{1}{c}{\textbf{Shape}} & \multicolumn{1}{c}{\boldmath{$Log(L)$}} & \multicolumn{1}{c}{\textbf{AIC}} \\
\midrule
\multicolumn{2}{l}{First stage}                  &                            &                           &                            &                          &                           &                            &                         \\
DJI                  & $-$0.20355 (0.215422)       & $-$0.15858   *** (0.042943)     & 0.97786   *** (0.023375)     & 0.24883 (0.221394)         & 0.86471   *** (0.068518)    & 5.43556  ** (2.795454)      & 1656.919                   & $-$6.383                 \\
FTSE                 & $-$0.10745   *** (0.00262)     & $-$0.14098   *** (0.028023)     & 0.988111   *** (0.000171)    & 0.08099  ** (0.037483)       & 0.86005   *** (0.054435)    & 5.56222   *** (1.331363)     & 1641.112                   & $-$6.334                 \\
GDAXI                & $-$0.05417   *** (0.004147)    & $-$0.18539   *** (0.028109)     & 0.994266  *** (0.000005)    & 0.03079 (0.022275)         & 0.86107   *** (0.046866)    & 3.66400  *** (0.650932)     & 1584.797                   & $-$6.108                 \\
GSPC                 & $-$0.26431  *** (0.068048)    & $-$0.13035   *** (0.035984)     & 0.971682  *** (0.007229)    & 0.26354   *** (0.064915)      & 0.77979   *** (0.055043)    & 6.07973  *** (1.761195)     & 1685.150                    & $-$6.485                 \\\midrule
\multicolumn{2}{l}{Second Stage}                 &                            &                           &                            &                          &                           &                            &                         \\
Joint                &                           & 0.05458   *** (0.010974)      & {0.88763   *** (0.025251)}              &                          &      & 6.13999   *** (0.555841)                    & 7507.884 & $-$28.941      \\
\bottomrule          
\end{tabular}%
}
\noindent\footnotesize{Note: GARCH models were estimated with the Student's $t$ distribution. Asymptotic standard errors are given in parentheses. Mean model---ARFIMA: DJI(1, 0, 2), FTSE(0, 0, 0), GDAXI(2, 0, 0), GSPC(4, 0, 1), respectively. We create GARCH estimates for the four stock market indices portfolios in the first phase. In the second stage, GARCH estimates of the joint series are evaluated while considering the conditional correlation among the four groups.   ** Significance at the 5\% level,   *** at the 1\% level.}
\end{table}\vspace{-15pt}

\begin{table}[H]
\caption{Results from multivariate EGARCH (DCC) model---Cryptocurrencies.}
\label{tab:my-table}
\renewcommand{\arraystretch}{1.25}
\resizebox{\textwidth}{!}{%
\begin{tabular}{lp{0.153\linewidth}p{0.15\linewidth}p{0.15\linewidth}p{0.14\linewidth}p{0.15\linewidth}p{0.15\linewidth}p{0.1\linewidth}p{0.1\linewidth}}
\toprule
\multicolumn{1}{c}{} & \multicolumn{1}{c}{\boldmath{$\omega$}} & \multicolumn{1}{c}{\boldmath{$\alpha_1$}} & \multicolumn{1}{c}{\boldmath{$\beta_1$}} & \multicolumn{1}{c}{\boldmath{$\gamma_1$}} & \multicolumn{1}{c}{\textbf{Skew}} & \multicolumn{1}{c}{\textbf{Shape}} & \multicolumn{1}{c}{\boldmath{$Log(L)$}} & \multicolumn{1}{c}{\textbf{AIC}} \\
\midrule
First Stage          &                           &                            &                           &                            &                          &                           &                            &                         \\
BTC                  & $-$0.02862   *** (0.007336)    & $-$0.089649  *** (0.020531)    & 0.994565  *** (0.00263)     & 0.165754 ** (0.056144)      & 1.01418  *** (0.038105)    & 2.332726  *** (0.065807)    & 1514.983                   & $-$4.123                  \\
ETH                  & $-$0.333921 ** (0.129402)    & $-$0.083583 * (0.033099)      & 0.943549  *** (0.021325)    & 0.164941 ** (0.059533)      & 0.988406  *** (0.046471)   & 2.906735  *** (0.335029)    & 1295.614                   & $-$3.520                  \\
XRP                  & $-$0.085255 ** (0.041778)    & $-$0.114611 ** (0.041343)     & 0.984135  *** (0.007442)    & 0.274431  *** (0.087779)     & 1.019635  *** (0.040798)   & 2.340565  *** (0.200588)    & 1397.988                   & $-$3.803                  \\
ADA                  & $-$0.896513 * (0.505929)     & $-$0.049195 (0.040657)       & 0.846718  *** (0.086361)    & 0.172597 * (0.070766)       & 1.039076  *** (0.050585)   & 3.545309  *** (0.511266)    & 1186.803                   & $-$3.222                  \\\midrule
\multicolumn{2}{l}{Second Stage}                 &                            &                           &                            &                          &                           &                            &                         \\
Joint                &                           & 0.024248 ** (0.010787)      & 0.936152  *** (0.025525)    &                            &                          & 4.000001  *** (0.170452)    & 6516.831                   & $-$17.729     \\
\bottomrule           
\end{tabular}%
}
\noindent\footnotesize{Note: GARCH models were estimated with the Student's $t$ distribution. Asymptotic standard errors are given in parentheses. Mean model---ARFIMA: BTC(0, 0, 1), ETH(0, 0, 2), XRP(1, 0, 0), ADA(0, 0, 2), respectively. We create GARCH estimates for the four cryptocurrencies in the first phase. In the second stage, GARCH estimates of the joint series are evaluated while considering the conditional correlation among the four groups.  * Significance at 10\% level,  ** at the 5\% level,   *** at the 1\% level.}
\end{table}

The multivariate DCC-GARCH parameter estimates for the most suitable EGARCH-type model are summarized in Table \ref{tab:dcc_stock}, which presents the performance of the study stock market indices. Empirical findings exhibit that all stock market indices except for the Dow Jones (DJI) index and the German DAX (GDAXI) index, which both show non-significant leverage impact parameter $\gamma_1$, have computed coefficients $\omega$, $\alpha_1$, $\beta_1$, $\gamma_1$, and {shape} 
 that are statistically significant. In addition, the Dow Jones (DJI) index reports a non-significant $\omega$ $-$0.20355, for the other three indices $\omega$ is significant to 0.001 level, for FTSE is $-$0.10745, GDAXI is $-$0.05417, and for S\&P500 is $-$0.26431, respectively. The findings show that lagged shocks and volatility significantly and positively impact the current conditional volatility. The values of the permanent parameters $\beta_1$ range from low 0.971682 for the S\&P500 (GSPC) index to high 0.994266 for the German DAX (GDAXI) index are close to one, positive, and significant. The German DAX index (GDAXI) has a leverage effect $\gamma_1$ parameter value of low 0.03079, while the S\&P500 index (GSPC) reports a value of 0.26354 high. Additionally, adjustments to previous shock parameters ($\alpha_1$) are negative ($-$0.15858 for the DJI, $-$0.14098 for the FTSE, $-$0.18539 for the GDAXI, and $-$0.13035 for the S\&P 500), respectively. Finally, all of the parameters for the leverage effect are positive, and the excess kurtosis parameter $shape$ yield values ranging from low 3.66400 for the German DAX index (GDAXI) to high 6.07973 for the S\&P500 index (GSPC), following the same pattern as the leverage effect $\gamma_1$ parameter.

Considering the previous discussion, GARCH models have been used to study the relationship between conditional variance and asset risk premium. The results show that volatility considerably rises in the wake of negative news. The parameter $\gamma_1$ is typically negative for the stock market, according to a study by \cite{nelson1991conditional} discussing the conditional heteroskedasticity in asset returns. All stock indices in this analysis had significant positive $\gamma_1$ values, which shows that the own-shock spillover has no effect on current volatility and that the daily fluctuations of these indexes rise after good news. The only exceptions are the DJI and the GDAXI indices, both positive but not statistically significant. It demonstrates that pandemic crashes may have significantly impacted these~indices. 

To improve the estimation accuracy, in the second stage, Table \ref{tab:dcc_stock}, we employed a multivariate EGARCH DCC(1, 1) model on a student's t-distribution to parameterize the distributions following the first stage with the EGARCH model and to improve the accuracy of the estimation \citep{engle2001theoretical}. The 41 parameters estimated by the DCC model are all statistically significant at the 1\% level. The presented joint estimates for $\alpha_1$ and $\beta_1$ are, respectively, 0.05458 and 0.88763. The multivariate joint shape is similarly substantial and moderately low, and the multivariate parameters are significantly positive (6.13999). This moderate to strong positive correlation between the stock market indices may affect their combined $\alpha_1$, leading to favorable reactions to early shocks.

The performance of the cryptocurrency returns for the most acceptable EGARCH-type model (First Stage) is shown in Table \ref{tab:my-table}, which also reports the multivariate DCC-GARCH parameter estimates (Second Stage). This empirical study revealed that cryptocurrency returns reported computed coefficients $\omega$, $\alpha_1$, $\beta_1$, $\gamma_1$, and $shape$ that are statistically significant except for Cardano (ADA) that report a non-significant $\alpha$ $-$0.049195. All cryptocurrency report significant $\omega$ to various statistically significant levels. For example, BTC is $-$0.02862 ($\alpha$ = 0.001), ETH is $-$0.333921 ($\alpha$ = 0.05), XRP is $-$0.085255 ($\alpha$ = 0.05), and ADA is $-$0.896513 ($\alpha$ = 0.01). The results demonstrate that lagged shocks and volatility have an extensive and favorable effect on the current conditional volatility. The values of the permanent parameters $\beta_1$ range from low 0.846718 for the Cardano (ADA) to high 0.994565 for the Bitcoin (BTC) and are close to one, positive, and significant ($\alpha$ = 0.001). Moreover, Ethereum (ETH) has a leverage effect $\gamma_1$ parameter value of 0.164941 low, while Ripple (XRP) provides a value of 0.274431 high. The adjustments to previous shock parameters ($\alpha_1$) are negative as expected with values as $-$0.089649 for BTC, $-$0.083583 for ETH, $-$0.114611 for the XRP, and $-$0.049195 for ADA, respectively. The excess kurtosis parameter $shape$ product values range from 2.332726 low for the BTC to 3.545309 high for ADA.

In the Second Stage, Table \ref{tab:my-table}, we used a multivariate EGARCH DCC(1, 1) model on a student's t-distribution to parameterize the distributions after the first stage using the EGARCH model and to increase the accuracy of the estimation \citep{engle2001theoretical}. The order of the DCC model estimated 37 parameters, all of which are statistically significant. The reported joint estimates for $\alpha_1$ and $\beta_1$ are 0.024248 and 0.936152, respectively. The multivariate parameters are significantly positive 4.000001, $\alpha$ = 0.001, and the multivariate joint shape is similarly significant and comparatively small. This moderately positive correlation may affect the aggregate $\alpha_1$ of the cryptocurrencies, resulting in positive responses to initial shocks. 

Moreover, we assessed the stock market indices portfolio's historical returns for the entire dataset sample ranging from 1 January  2019 to 31 December  2020. We employ a VaR model, and a four-moment modified VaR model using the Cornish--Fischer (CFVaR) expansion for three confidence levels, i.e., 90\%, 95\%, and 99\%, following a common approach of measuring the downside risk of a portfolio \citep{conlon2020safe, ali2021downside}. The findings in Table \ref{tab:my-VAR_stock} show that for Value-at-Risk (VaR) at a 95\% confidence interval, the most significant loss is a 2.4\% loss (DJI, S\&P500); however, all stock market indices show a similar comparative loss, with tiny differences. At the 99\% confidence level, DJI offers the highest risk, followed again by the S\&P500 and the GDAXI, with risk high at 5.7\%, 4.8\%, and 4.3\%, respectively. The higher VaR is a result of the lower confidence level. A higher confidence level will result in a more significant loss percentage, and a decreased confidence level might lead to a lower rate of failure; this is reasonable and assists in understanding how the VaR operates. Such results were expected if we consider that the Dow (DJI) index crashed during March 2020 financial market black days \citep{forbes2021}. However, the financial markets unexpectedly recovered most of their value within a month to a surge that pushed the stock indices back to the pre-COVID period. For the same period, CFVaR at a 99\% confidence level reports the worst loss of 9.6\% (DJI), followed by the S\&P500 index at a high of 8.5\% with the portfolio, and the FTSE index reports (6.9\%). As such, Table \ref{tab:my-VAR_stock} shows that all results are higher than the VaR results. The Cornish--Fisher VaR (CFVaR) will provide a higher loss estimate than the typical VaR when the returns are negatively skewed (Table \ref{summarystock}), which is exhibited as such in this study.

\begin{table}[H]
\caption{\textls[-25]{Value at Risk (VaR) and Cornish--Fisher expansion (CFVaR) estimations---Stock market indices.}}
\label{tab:my-VAR_stock}
\renewcommand{\arraystretch}{0.99}
\setlength{\tabcolsep}{10pt}
\resizebox{\textwidth}{!}{%
\begin{tabular}{llrrrrrr}
\toprule
       &                       & \multicolumn{2}{l}{\textbf{(January 2019--December 2020)}}               & \multicolumn{2}{l}{\textbf{(January--December 2019)}}               & \multicolumn{2}{l}{\textbf{(January--December 2020)}}               \\ \cmidrule{2-8}
\multirow{-2}{*}{\textbf{Indices}\vspace{3pt}} & \multicolumn{1}{c}{\textbf{\emph{p}}} & \multicolumn{1}{c}{\textbf{VaR}} & \multicolumn{1}{c}{\textbf{CFVaR}} & \multicolumn{1}{c}{\textbf{VaR}} & \multicolumn{1}{c}{\textbf{CFVaR}} & \multicolumn{1}{c}{\textbf{VaR}} & \multicolumn{1}{c}{\textbf{CFVaR}} \\
\midrule
DJI                                         & 90\%                  & 0.013                   & 0.007                     & 0.007                   & 0.010                     & 0.020                   & 0.021                     \\
                                            & 95\%                  & 0.024                   & 0.026                     & 0.012                   & 0.013                     & 0.032                   & 0.037                     \\
                                            & 99\%                  & 0.057                   & 0.096                     & 0.026                   & 0.020                     & 0.073                   & 0.089  \\
\midrule
FTSE                                        & 90\%                  & 0.013                   & 0.010                     & 0.008                   & 0.010                     & 0.018                   & 0.020                     \\
                                            & 95\%                  & 0.022                   & 0.024                     & 0.010                   & 0.012                     & 0.033                   & 0.033                     \\
                                            & 99\%                  & 0.040                   & 0.069                     & 0.022                   & 0.017                     & 0.046                   & 0.068                     \\\midrule
GDAXI                                       & 90\%                  & 0.014                   & 0.009                     & 0.009                   & 0.011                     & 0.020                   & 0.020                     \\
                                            & 95\%                  & 0.023                   & 0.025                     & 0.016                   & 0.014                     & 0.037                   & 0.034                     \\
                                            & 99\%                  & 0.043                   & 0.079                     & 0.024                   & 0.019                     & 0.054                   & 0.078                     \\\midrule
GSPC                                        & 90\%                  & 0.012                   & 0.008                     & 0.007                   & 0.009                     & 0.018                   & 0.020                     \\
                                            & 95\%                  & 0.024                   & 0.025                     & 0.012                   & 0.013                     & 0.034                   & 0.035                     \\
                                            & 99\%                  & 0.048                   & 0.085                     & 0.025                   & 0.020                     & 0.066                   & 0.081     \\
\bottomrule               
\end{tabular}%
}
\noindent\footnotesize{Note: The Value-at-Risk (VaR) and the four-moment modified Value-at-Risk (CFVaR) using the Cornish--Fisher expansion are reported in this table.}
\end{table}\vspace{-6pt}

During the COVID-19 outbreak period (January--December 2020), the Dow (DJI) index reports VaR's highest loss of 7.3\%, followed by the S\&P500 (6.6\%), GDAXI with 5.4\%, and the FTSE with 4.6\%, respectively, similar to the entire period dataset. Next, we assess the period with the CFVaR approach (see Table \ref{tab:my-VAR_stock}). Our results show a similar loss direction, with Dow (DJI) reporting the highest loss of 8.9\%, followed by the S\&P500 (8.1\%) index. The highest difference between the VaR and the CFVaR estimation is 2.4\% in the GDAXI index, and the average difference is 1.9\%. The results are not unexpected. A more significant CFVaR estimate than the typical VaR will be provided if the returns are not normally distributed, which is the case in this study. The results align with the study by \cite{MAZUR2021101690} found that extremely asymmetric volatility is present in underperforming stocks and negatively correlates with stock market returns. In this sense, businesses responded in various ways during the COVID-19 outbreak income shock.

Table \ref{tab:VaR_crypto} presents the confidence level of the historical returns of the cryptocurrencies market portfolio. Accordingly, at the 90\% confidence level for the value-at-risk (VaR) during the entire period, we encounter the worst loss with Cardano (ADA) of 6.1\%, followed by Ethereum (ETH) with a loss of 5.2\%, which is consistent with the correlation between them. Bitcoin reports less estimated risk (4.2\%) than the remaining cryptocurrencies, followed by Ripple (XRP) (4.4\%). This trend continues, with BTC reporting a smaller loss at 99\% confidence, at 9.1\%, followed by ETH with 11.4\%. For the same period, the CFVaR generated contradictory results. At the confidence level of 90\%, all cryptocurrencies except Ethereum (ETH) yielded negative results, i.e., BTC ($-$1.0\%), ADA ($-$4.0\%), and XRP ($-$9.2\%). It would be implied by the negative CFVaR that there is a significant likelihood of a profit for the portfolio. Ethereum (ETH) is the only cryptocurrency to report positive values at every confidence level. Because of the correlation between BTC and ETH, we would expect something similar here. As opposed to that, ETH reports the lowest loss of 9.6\% at the 99\% confidence level, followed by BTC. The XRP and the ADA report the highest loss of 31.9\%, and 25.6\%, respectively. A reason might be that both cryptocurrencies show a robust correlation.

During the COVID-19 outbreak period (January--December 2020), the VaR results in Table \ref{tab:VaR_crypto} show that at 90\% confidence level, BTC generated the lowest loss (4.1\%) followed by XRP (4.9\%). At a 99\% confidence level, ADA reported the worst loss (14.9\%) followed by XRP (14.7\%) loss. BTC reported the lowest loss (8.4\%), which is significantly lower than the ADA. Similar to the entire period of the study, as well as during the COVID-19 pandemic, there is a robust correlation between ADA and XRP. In Table \ref{tab:VaR_crypto}, the CFVaR during the pandemic reports negative results for the entire cryptocurrencies portfolio except the Ethereum (ETH) for confidence levels of 90\% and 95\%. For the 99\% confidence level, ADA and XRP show similar risk at 21.1\%, followed by BTC with 13.9\% loss. Ethereum (ETH) reports less risky results at a 9.7\% loss. The maximum difference between the cryptocurrencies loss is 11.4\%, ADA and XRP to ETH at the 99\% confidence level. These results contradict the risk reported in this study for the pre-pandemic period (January--December  2019), where Bitcoin (BTC) showed a higher risk value noting at the 99\% confidence level of 11.9\% risk, higher than the remaining cryptocurrencies. During the pandemic, BTC's trading behavior was more stable. {The results confirm the study by} \cite{yan2022garch}, {which found that COVID-19 positively affected the returns of cryptocurrencies, and the varying correlations were robust.}

\begin{table}[H]
\caption{Value at Risk (VaR) and Cornish--Fisher expansion (CFVaR) estimations---Cryptocurrencies.}
\label{tab:VaR_crypto}
\renewcommand{\arraystretch}{1.05}
\setlength{\tabcolsep}{9pt}
\resizebox{\textwidth}{!}{%
\begin{tabular}{llrrrrrr}
\toprule
       & \textbf{Period:}                       & \multicolumn{2}{l}{\textbf{(January 2019-December 2020)}}               & \multicolumn{2}{l}{\textbf{(January--December 2019)}}               & \multicolumn{2}{l}{\textbf{(January--December 2020)}}               \\  \cmidrule{2-8}
\multirow{-2}{*}{\textbf{Crypto}\vspace{3pt}} & \multicolumn{1}{c}{\emph{\textbf{p}}} & \multicolumn{1}{c}{\textbf{VaR}} & \multicolumn{1}{c}{\textbf{CFVaR}} & \multicolumn{1}{c}{\textbf{VaR}} & \multicolumn{1}{c}{\textbf{CFVaR}} & \multicolumn{1}{c}{\textbf{VaR}} & \multicolumn{1}{c}{\textbf{CFVaR}} \\ \midrule
BTC    & 90\%                  & 0.042                   & $-$0.010                    & 0.042                   & 0.044                     & 0.041                   & $-$0.043                    \\
       & 95\%                  & 0.055                   & 0.025                     & 0.057                   & 0.064                     & 0.054                   & $-$0.018                    \\
       & 99\%                  & 0.091                   & 0.189                     & 0.094                   & 0.119                     & 0.084                   & 0.139                     \\\midrule
ETH    & 90\%                  & 0.052                   & 0.048                     & 0.045                   & 0.053                     & 0.055                   & 0.042                     \\
       & 95\%                  & 0.071                   & 0.061                     & 0.076                   & 0.066                     & 0.070                   & 0.055                     \\
       & 99\%                  & 0.114                   & 0.096                     & 0.110                   & 0.090                     & 0.113                   & 0.097                     \\\midrule
XRP    & 90\%                  & 0.044                   & $-$0.092                    & 0.039                   & 0.045                     & 0.049                   & $-$0.072                    \\
       & 95\%                  & 0.066                   & $-$0.028                    & 0.060                   & 0.062                     & 0.076                   & $-$0.034                    \\
       & 99\%                  & 0.119                   & 0.319                     & 0.106                   & 0.100                     & 0.147                   & 0.211                     \\\midrule
ADA    & 90\%                  & 0.061                   & $-$0.040                    & 0.057                   & 0.057                     & 0.068                   & $-$0.050                    \\
       & 95\%                  & 0.084                   & 0.008                     & 0.077                   & 0.069                     & 0.093                   & $-$0.013                    \\
       & 99\%                  & 0.142                   & 0.256                     & 0.102                   & 0.087                     & 0.149                   & 0.211                    \\ \bottomrule
\end{tabular}%
}
\noindent\footnotesize{Note: The Value-at-Risk (VaR) and the four-moment modified Value-at-Risk (CFVaR) using the Cornish--Fisher expansion are reported in this table.}
\end{table}

\begin{figure}[H]

\begin{adjustwidth}{-\extralength}{0cm}
\centering 
    \subfloat{\includegraphics[width=0.595\textwidth]{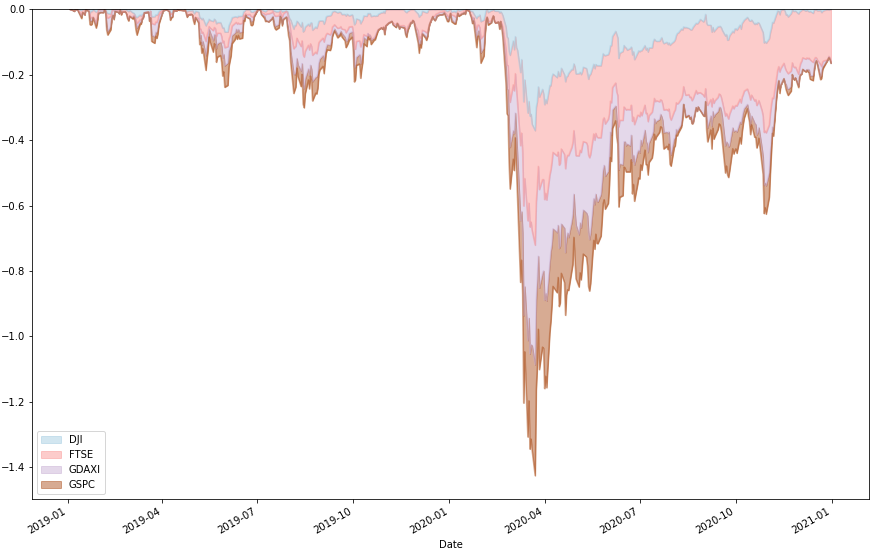}\label{fig:f1}}
  \hspace{0.5mm}
  \subfloat{\includegraphics[width=0.553\textwidth]{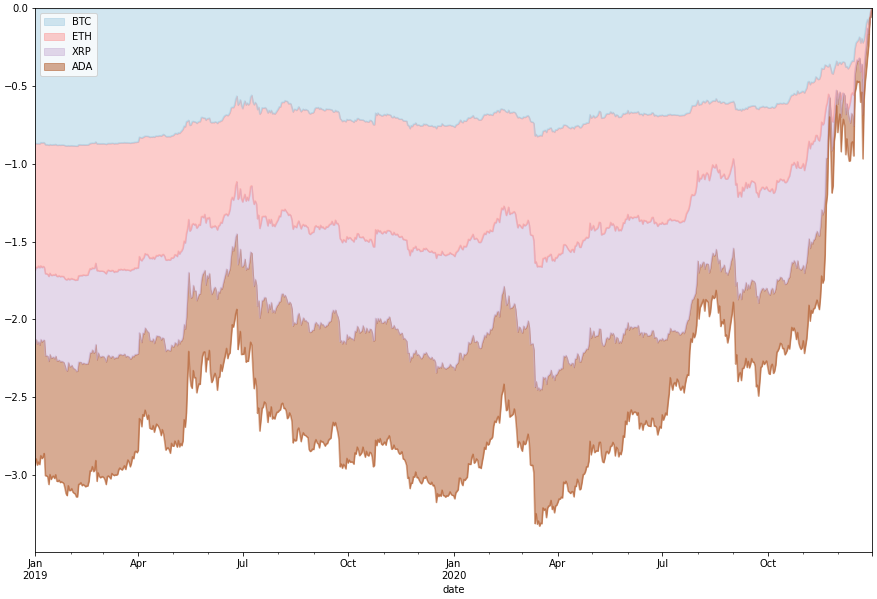}\label{fig:f2}}
\end{adjustwidth}
  \caption{Financial Market's drawdown during January 2019--December 2020.{(\textbf{a}) Stock market returns drawdown. (\textbf{b}) Crypto market returns drawdown.} 
}
  \label{fig:downside}
\end{figure}

Figure \ref{fig:downside} provides the downside risk of both portfolios' during the entire period of the study. It is an evaluation of how much funds could be lost due to a security's ability to lose value in the event that market conditions change. It is employed to comprehend the worst-case event of asset investment. In Figure \ref{fig:downside}a, we encounter a sharp drawdown that recovers relatively quickly. In Figure \ref{fig:downside}b, the cryptocurrencies portfolio experiences a longer modest drawdown that lasts for a long time, sometimes less hurting than a sharp one. 

\section{Conclusions}\label{sec6}
Considering the volatility of any financial market portfolio involves the evaluation of the model's characteristics. This research focuses on evaluating the volatility dynamics of financial asset returns on four global stock market indices and four leading cryptocurrencies during the COVID-19 outbreak utilizing a two-stage multivariate DCC-GARCH Student-t distribution model. During the pandemic, the stock market indices collapsed in March 2020, leading to one of the most significant stock market crashes showing a decline of close to 40\%. For example, the Dow Industrial Index (DJI) lost 37\% of its value in four trading days \citep{forbes2021} due to the pandemic and public administration decisions.

Investors should be mindful while making investment decisions in financial assets. We observed that the assets portfolio's current conditional variance was significantly affected by its historical volatility and shocks. Our findings indicate that the COVID-19 pandemic has improved the underlying assets' broad correlation. This work contributes by initially extending and verifying earlier GARCH models using EGARCH terms, an extension of the GARCH-family model to predict and anticipate volatility. Moreover, the research incorporated pairwise models between the assets portfolios estimated using the multivariate DCC-EGARCH model to produce more robust estimations. The DCC-EGARCH model also offers the benefit of including the analysis of dynamic beta values. 

We quantified the relative risk using two widely used metrics of downside risk, portfolio value-at-risk (VaR) and value-at-risk (CFVaR), based on the Cornish--Fisher expansion, an approach suited for incorporating higher-order distributional properties associated with drastic price changes. According to the accuracy estimation, the VaR performance comparison results with the assets portfolios differ. For example, during the COVID-19 outbreak, the Dow (DJI) index reports VaR's highest loss, followed by the S\&P500. Similarly, a close loss direction is informed by assessing the stock market using the CFVaR approach, with Dow (DJI) showing the highest loss, followed again by the S\&P500 index. Conversely, during the pandemic, the CFVaR reports negative risk results for the entire cryptocurrency portfolio except the Ethereum (ETH) for confidence levels of 90\% and 95\%. As such, investment managers should choose GARCH-type models with a long memory to estimate the VaR of the portfolios, considering the high volatility dynamics observed in all financial assets. 

Future research topics might examine the effects of earlier occurrences resembling COVID-19, how COVID-19 may differ from those earlier events, and obtain an optimal portfolio in a highly dependent volatile financial market environment. In addition, investors and portfolio managers actively investing in financial assets will find our observations of considerable interest. Overall, our findings offer information to regulators and investors on risk management and optimal asset allocation. Investors can use optimal portfolios to create portfolios that decrease risk exposure during and after a crisis. However, if authorities wish to prevent negative repercussions from infectious shocks, they must closely monitor changes in the financial assets and follow up with caution. 

\vspace{6pt}

\funding{This research received no external funding.}

\dataavailability{Data are publicly available.}

\conflictsofinterest{The authors declare no conflict of interest.}

\begin{adjustwidth}{-\extralength}{0cm}
\printendnotes[custom]
\reftitle{References}

\PublishersNote{}

\end{adjustwidth}
\end{document}